\documentclass[pra,aps,amsfonts,amsmath,superscriptaddress, twocolumn,showpacs,floatfix,nofootinbib]{revtex4-1}
\usepackage{amsfonts}
\usepackage{amsmath}
\usepackage{amssymb}
\usepackage{bm}
\usepackage{graphicx}
\usepackage{hyperref}
\hypersetup{
     colorlinks,
     linkcolor={red!50!black},
     citecolor={blue!50!black},
     urlcolor={blue!80!black}
}
\usepackage{textcomp}
\usepackage[table]{xcolor}
\usepackage{multirow}

\usepackage{array}
\newcolumntype{M}[1]{>{\centering\arraybackslash}m{#1}}

\begin{document}

\title{\bf Effective proton-neutron interaction near the drip line from unbound states in $^{25,26}$F}

\author{M.~Vandebrouck} 
\email{Present address: Irfu, CEA, Universit\'e Paris-Saclay, 91191 Gif-sur-Yvette, France.
E-mail address: marine.vandebrouck@cea.fr}
\address{Grand Acc\'el\'erateur National d'Ions Lourds (GANIL),
CEA/DSM-CNRS/IN2P3, Bvd Henri Becquerel, 14076 Caen, France}
\author{A.~Lepailleur} 
\address{Grand Acc\'el\'erateur National d'Ions Lourds (GANIL),
CEA/DSM-CNRS/IN2P3, Bvd Henri Becquerel, 14076 Caen, France}
\author{O.~Sorlin} 
\address{Grand Acc\'el\'erateur National d'Ions Lourds (GANIL),
CEA/DSM-CNRS/IN2P3, Bvd Henri Becquerel, 14076 Caen, France}

\author{T.~Aumann}
\affiliation{Institut f\"ur Kernphysik, Technische Universit\"at Darmstadt, 64289 Darmstadt, Germany}
\affiliation{GSI Helmholtzzentrum f\"ur Schwerionenforschung, 64291 Darmstadt, Germany}
\author{C.~Caesar}
\affiliation{Institut f\"ur Kernphysik, Technische Universit\"at Darmstadt, 64289 Darmstadt, Germany}
\affiliation{GSI Helmholtzzentrum f\"ur Schwerionenforschung, 64291 Darmstadt, Germany}
\author{M.~Holl}
\affiliation{Institut f\"ur Kernphysik, Technische Universit\"at Darmstadt, 64289 Darmstadt, Germany} 
\author{V.~Panin}
\affiliation{Institut f\"ur Kernphysik, Technische Universit\"at Darmstadt, 64289 Darmstadt, Germany} 
\author{F.~Wamers}
\affiliation{Institut f\"ur Kernphysik, Technische Universit\"at Darmstadt, 64289 Darmstadt, Germany}
\affiliation{GSI Helmholtzzentrum f\"ur Schwerionenforschung, 64291 Darmstadt, Germany}

\author{S.\ R.\ Stroberg}
\affiliation{TRIUMF, 4004 Wesbrook Mall, Vancouver, British Columbia, 
V6T 2A3, Canada}
\author{J.\ D.\ Holt}
\affiliation{TRIUMF, 4004 Wesbrook Mall, Vancouver, British Columbia, 
V6T 2A3, Canada}
% ...

\collaboration{R$^3$B collaboration}

\begin{abstract}
\textbf{Background:} Odd-odd nuclei, around doubly closed shells, have been extensively used to study proton-neutron interactions. However, the evolution of these interactions as a function of the binding energy, ultimately when nuclei become unbound, is poorly known. The $^{26}$F nucleus, composed of a deeply bound $\pi0d_{5/2}$ proton and an unbound $\nu0d_{3/2}$ neutron on top of an $^{24}$O core, is particularly adapted for this purpose. The coupling of this proton and neutron results in a $J^{\pi} = 1^{+}_1 - 4^{+}_1$ multiplet, whose energies must be determined to study the influence of the proximity of the continuum on the corresponding proton-neutron interaction. The $J^{\pi} = 1^{+}_1, 2^{+}_1,4^{+}_1$ bound states have been determined, and only a clear identification of the $J^{\pi} =3^{+}_1$ is missing.

\textbf{Purpose:} We wish to complete the study of the $J^{\pi} = 1^{+}_1 - 4^{+}_1$ multiplet in $^{26}$F, by studying the energy and width of the $J^{\pi} =3^{+}_1$ unbound state. The method was firstly validated by the study of unbound states in $^{25}$F, for which resonances were already observed in a previous experiment.

\textbf{Method:} Radioactive beams of $^{26}$Ne and $^{27}$Ne, produced at about $440A$\,MeV by the FRagment Separator at the GSI facility, were used to populate unbound states in $^{25}$F and $^{26}$F via one-proton knockout reactions on a CH$_2$ target, located at the object focal point of the R$^3$B/LAND setup. The detection of emitted $\gamma$-rays and neutrons, added to the reconstruction of the momentum vector of the  $A-1$ nuclei, allowed the determination of the energy of three unbound states in $^{25}$F and two in $^{26}$F. 

\textbf{Results:} Based on its width and decay properties, the first unbound state in $^{25}$F is proposed to be a $J^{\pi} = 1/2^-$ arising from a $p_{1/2}$ proton-hole state. In $^{26}$F, the first resonance at 323(33)~keV is proposed to be the $J^{\pi} =3^{+}_1$ member of the $J^{\pi} = 1^{+}_1 - 4^{+}_1$ multiplet. Energies of observed states in $^{25,26}$F have been compared to calculations using the independent-particle shell model, a phenomenological shell-model,  and the ab initio valence-space in-medium similarity renormalization group method.

\textbf{Conclusions:} The deduced effective proton-neutron interaction is weakened by about 30-40\% in comparison to the models, pointing to the need of implementing the role of the continuum in theoretical descriptions, or to a wrong determination of the atomic mass of $^{26}$F.
\end{abstract}

\maketitle

\section{Introduction}

The study of odd-odd nuclei is experimentally challenging, as such systems display many states of angular momentum $J$ built from the coupling of the odd proton $j_p$ and neutron $j_n$, leading to  $|j_p-j_n| \leq J \leq |j_p+j_n|$ multiplets. Moreover, long-lived isomers are often present when states of extreme  $|j_p-j_n|$ and $|j_p+j_n|$ values lie close in energy, and different experimental techniques may be required to determine the energy E($J$) of all states in a given multiplet. Such studies on odd-odd nuclei close to doubly magic ones, however, are rewarded by the wealth of information obtained on proton-neutron interactions \cite{Schiffer1976}, when an independent-particle shell model (IPSM) scheme is used. 
For the members of a given multiplet, the experimental energies E($J$) of the states are empirically observed to vary parabolically as a function of $J(J+1)$ \cite{Paar1979}. 
These E($J$) are used to determine the proton-neutron interactions, Int($J$), derived from a shift of E($J$), in order to obtain Int$(J)=0$ when the proton and neutron added to the closed shells do not interact (\cite{Lepailleur2013} and Sect.~\ref{par:26FDiscussion} of the present work). It follows that a parabolic law can be applied to Int($J$) as a function of $J(J+1)$ as well. When interpreted in terms of a low-order multipole expansion, the monopole part, which is the $(2J\!+\!1)$-weighted average of Int$(J)$, contains information on the strength of the nuclear interaction. The dominant quadrupole part, which depends in principle on the relative orientation between the interacting valence proton and neutron only, breaks the degeneracy between multiplet levels and generate the observed parabolic behavior \cite{Casten2000}. This simple picture restricts to nuclei near closed shells as it neglects effects of the coupling to other bound or unbound states of similar $J^{\pi}$ values, that can modify the shape of the parabola. 

Further complications to this simple model arise for nuclei near the drip lines, where some (if not all) states comprising multiplets become unbound. Besides the fact that their characterization (i.e. energy, width, orbital angular momentum $\ell$) is less certain than for bound states, unbound states with pure configurations exhibit large widths, due to their large overlap with states in the $(A-1)$ nucleus. Resonances are expected to broaden as their energy increases, leading progressively to a continuum of indistinguishable, overlapping resonances. Deviations to this global trend occur when unbound states are trapped in the nuclear potential by high centrifugal barriers, or have a very poor configuration overlap with the available decay channels (see e.g.\cite{DeGrancey2016}).

Though challenging, the extension of these experimental investigations to the drip-line regions would provide new information on the behavior of Int($J$) in extreme proton-neutron asymmetries and when one or more states of the multiplet are unbound. The validity of a bound single-particle approach to drip-line nuclei is of interest for the study of drip-line phenomena such as nuclear halos, islands of inversion, and in nuclear astrophysics for the modeling of neutron stars.

Two recent studies provided some first insights into these questions. The comparison of the two odd-odd mirror nuclei $^{16}$N and $^{16}$F, the first being bound, the second being proton-unbound, both having rather pure single-particle configuration, showed an orbital- and binding-energy-dependent reduction of the experimental proton-neutron interaction (monopole part) of up to 40\% between the two mirror nuclei. This effect was attributed to the large radial extension of certain orbits that probe the continuum \cite{Stefan2014}. Studies of the $N\!=\!17$ odd-odd isotones towards the neutron drip line (from $Z\!=\!13$ to $Z\!=\!9$) have suggested, making use of a tentative assignment of the unbound $J\!=\!3$ state in $^{26}$F \cite{Lepailleur2013,Frank2011,Basunia2016}, a gradual reduction of the experimental proton-neutron interaction with increasing neutron-to-proton asymmetry \cite{Lepailleur2015}, rather than an abrupt change at the drip line. 

As discussed in Ref.~\cite{Lepailleur2013}, the weakly bound $^{26}$F is one of the few ideal nuclei where we can study the impact of continuum effects on Int($J$). Lying close to the doubly magic $^{24}$O \cite{Hoffman2009, Kanungo2009, Tshoo2012}, whose first excited states lie above 4 MeV \cite{Hoffman2009,Tshoo2012}, low-energy states in $^{26}$F are, in the IPSM picture, expected to arise from the coupling of a deeply bound $\pi0d_{5/2}$ proton ($S_{p}(^{25}\mathrm{F}) \!=\! 14.43(14)$~MeV \cite{Wang2012}) with an unbound $\nu0d_{3/2}$ neutron ($S_{n}(^{25}\mathrm{O}) \!=\!  -749(10)$~keV \cite{Kondo2016,Caesar2013,Hoffman2008}). This $(\pi0d_{5/2})^1(\nu0d_{3/2})^1$ coupling results in a $J^{\pi} \!=\! 1^{+}_1 \!-\! 4^{+}_1$ multiplet (Fig.~\ref{fig:LevelScheme26F}(a)).
Energies of the bound $J^{\pi} \!=\! 1^{+}_1,  2^{+}_1$ and $4^{+}_1$ states were measured using different experimental techniques \cite{Stanoiu2012,Lepailleur2013,Jurado2007}, and only a firm 
identification of the $J^{\pi} = 3^{+}_1$ component is missing. In particular, the spin assignments of the ground state (1$^+$) \cite{Reed1999,Lepailleur2013}, and of the weakly bound  isomeric state (4$^+_1$ at 643~keV) \cite{Lepailleur2013}, were proposed from their decay pattern to low and high energy spin values, respectively, in the daughter nucleus $^{26}$Ne\footnote{The structure of the ground state of $^{26}$F was also investigated by the one-neutron knockout reaction at relativistic energies \cite{Carme2010}. The narrow inclusive momentum distribution of the $^{25}$F residue pointed to the presence of valence neutrons in the $1s_{1/2}$ state, in apparent contradiction with the neutron $0d_{3/2}$ configuration of the 1$^+$ ground state proposed above. An exclusive one-neutron knockout experiment is needed to isolate the contributions leading to $^{25}$F in either ground state or excited states, that would correspond to the knock-out from the last occupied or more deeply bound neutron orbitals, respectively.  We moreover point out that the one-neutron knockout reaction may have occurred from the, at that time unknown, weakly bound $4^+$ 2~ms-isomer. In such a case, the resulting one-neutron knockout momentum distribution from a weakly bound $0d_{3/2}$ orbit may be mimicking the one corresponding to an $1s_{1/2}$ orbit.}. A resonance was observed at $271(37)$~keV above the neutron threshold using the nucleon-exchange reaction \mbox{$^{26}$Ne $\rightarrow$ $^{26}$F} \cite{Frank2011}. However no spin assignment was proposed. The next likely 
multiplet would arise from the $(\pi0d_{5/2})^1(\nu1s_{1/2})^{-1}(\nu0d_{3/2})^2$ configuration, leading to $J^{\pi} = 2^{+},  3^{+}$ states (Fig.~\ref{fig:LevelScheme26F}(b)). In the case of single-particle
proton excitations, $J^{\pi} = 1^{+},  2^{+}$ states are formed by the $(\pi1s_{1/2})^1(\nu0d_{3/2})^1$ configuration (Fig.~\ref{fig:LevelScheme26F}(c)). None of these states has yet been observed.

In this article we have studied unbound states in $^{26}$F produced by the one-proton knockout reaction at the GSI facility. The knockout of a $0d_{5/2}$ proton 
from $^{27}$Ne should leave the $^{26}$F nucleus in the
$(\pi0d_{5/2})^1(\nu0d_{3/2})^1$ configuration (Fig.~\ref{fig:LevelScheme26F}(a)) and favor the production of  the 1$^+ - 4^+$ multiplet of states, including the 3$^{+}$. The $^{25}$F nucleus, also produced by one-proton knockout reaction from $^{26}$Ne, has been studied as well.  In both cases, results are compared  to previous experimental values.  

To gauge the validity of the IPSM nature of these multiplets, we compare to predictions of other theoretical models described in Section III: the phenomenological shell-model, which implicitly contains some aspects of continuum physics and three-nucleon (3N) forces, and the ab initio valence-space in-medium 
similarity renormalization group (IM-SRG) \cite{Tsuk12SM,Bogn14SM,Stroberg2015,Stroberg2016} based on two-nucleon (NN) and 3N forces, but neglecting the influence of the continuum.

\section{Experimental setup}

A  stable beam of $^{40}$Ar  was accelerated by the linear accelerator UNILAC and by the synchrotron SIS-18 at the GSI facility to an energy of $490A$\,MeV and impinged on a 4~g/cm$^2$-thick $^9$Be  target to induce fragmentation reactions, in which the $^{27}$Ne and $^{26}$Ne nuclei were produced. They  were  subsequently selected by the FRagment Separator  (FRS) \cite{Geissel1992}, whose magnetic rigidity was set to 9.05~Tm in order to favor the transmission of nuclei with $A/Z\!\sim\!2.7$. These secondary nuclei were transmitted to the R$^3$B/LAND experimental setup \cite{Aumann2007}, where they were identified on an event-by-event basis using \textit{i)} their energy-loss in the Position Sensitive silicon Pin diode (PSP) detector, \textit{ii)} their time of flight measured between two plastic scintillators, one located at the end of the FRS beam line, and the other (\textit{start} detector POS) placed a few meters before a 922 mg/cm$^2$ CH$_2$ reaction target. A total of $2.5 \times 10^5$ ($3.8 \times 10^5$) nuclei of $^{27}$Ne ($^{26}$Ne) impinged on the CH$_2$ target, with an energy at the entrance of $432$ $(456)A$\,MeV. 

This secondary target was surrounded by the 159 NaI crystals of the $4\pi$ Crystall Ball detector \cite{Metag1983}, each having a length of 20~cm and covering a solid angle of $\simeq 77$~msr. It allowed the detection of photons from excited fragments decaying in-flight and recoil protons at angles larger than $\pm 7$\textdegree\ in the laboratory frame. Each crystal was equipped with phototubes having a gain adapted for the detections of photons. In addition, the photomultipliers of the 64 most forward crystals had a second lower-gain readout, for the detection of recoil protons originating from knockout reactions. Two pairs of double-sided silicon strip detectors (DSSSD), with active areas of $72 \times 41$~mm$^2$ and strips 300~$\mu$m thick (110~$\mu$m pitch), were placed before and after the reaction target to determine the energy loss and to track the incoming and outgoing nuclei, e.g. $^{27}$Ne and $^{26}$F, respectively, in the case of a one-proton knockout reaction from $^{27}$Ne populating bound states in $^{26}$F. 

\begin{figure}[htb]
\begin{center}
\includegraphics[width=8.7 cm]{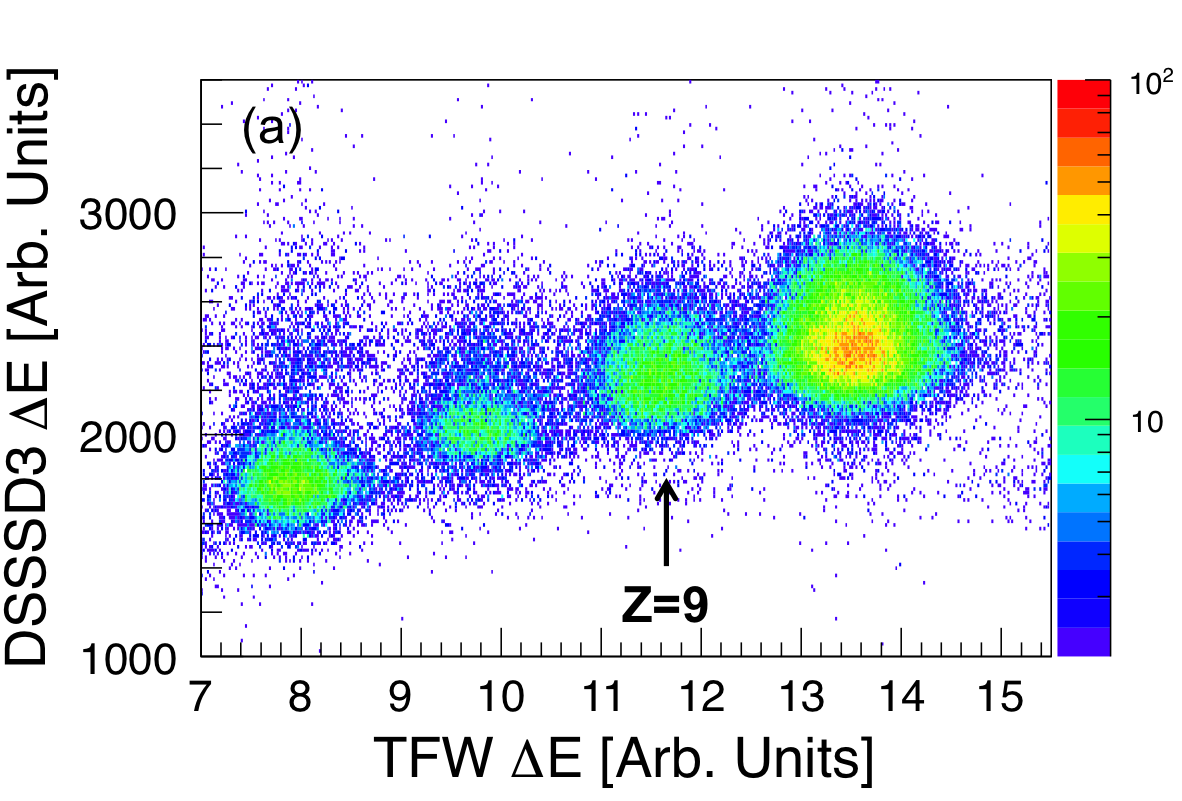}
\includegraphics[width=8.7 cm]{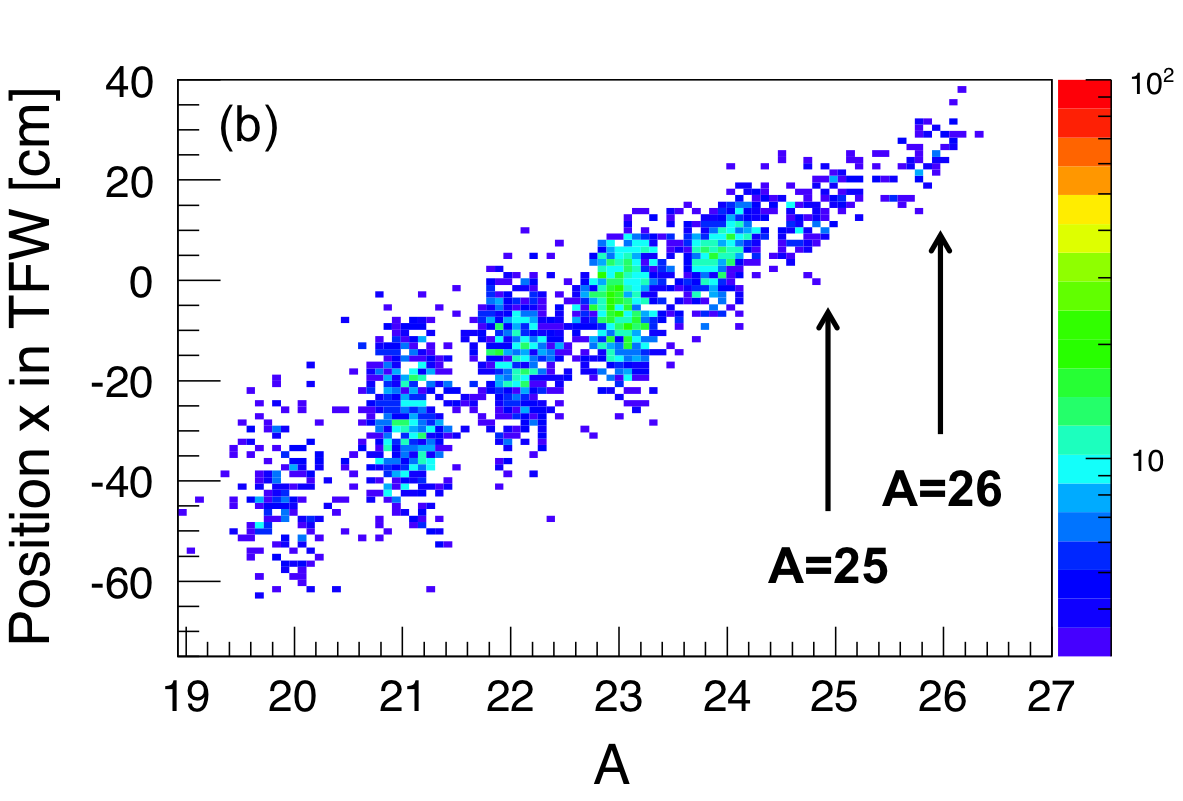}
\end{center} 
\caption{(Color online) (a) $Z\!=\!9$ transmitted nuclei obtained from the energy losses measured in the DSSSD located just after the target, and in the TFW. (b) Mass number $A$ identification of the transmitted $Z\!=\!9$ nuclei is obtained from their reconstructed trajectory in the dispersive plane, and from their time of flight (see text for details).
\label{fig:ID}}
\end{figure}

After having passed the downstream pair of DSSSDs, nuclei were deflected in the large dipole magnet ALADIN. Their horizontal position was measured  at the dispersive plane of ALADIN in two scintillating fiber detectors (GFIs), each composed of 480  fibers, covering a total active area of 50 $\times$ 50~cm$^2$. Their energy loss, position and time-of-flight were determined based on the information provided by the time-of-flight wall (TFW) placed 523~cm behind the last GFI. The TFW is composed of plastic scintillator paddles, 14 horizontal ones in the first plane and 18 vertical ones in the second plane, read-out on both sides by photomultipliers. The atomic number $Z$ of the transmitted nuclei was obtained from the determination of their energy losses in the DSSSD, placed after the target, and in the TFW. The mass ($A$) identification is obtained from the combined position information of the fragments in the DSSSD placed after the target, in the two GFIs and in the TFW, and from their velocity $\beta$, which was deduced from their time-of-flight between the POS detector and the TFW \cite{LepailleurThesis,CaesarThesis}. The identification plots of the reacted nuclei in ($Z$, $A$) obtained from all these pieces of information  are shown in Fig.~\ref{fig:ID}(a) and \ref{fig:ID}(b).

When  produced in an unbound state during the proton knockout reaction, nuclei may emit neutrons that are detected in the forward direction using the large area neutron detector LAND \cite{Blaich1992}. It is composed of 10 planes with 20 paddles each, placed alternatively in horizontal and vertical directions, each paddle covering an area of $200 \times 10$~cm$^2$ and having a thickness of 10~cm. Each paddle is made of 11 iron and 10 scintillator sandwiched sheets, so that, when a neutron interacts with iron nuclei, secondary protons are produced and detected with the plastic scintillators. A specific algorithm is used to reconstruct the hit profiles in LAND, and obtain from them the position of the first neutron-LAND interaction (with a spatial resolution of 5~cm FWHM) and the neutron time-of-flight  (with a resolution of 370~ps FWHM). The LAND detector was positioned 13~m downstream of the reaction target, covering forward angles of $\pm 79$~mrad. The intrinsic efficiency for a $\sim$450~MeV neutron is about 60\%, and the geometric acceptance is 100\% up to a fragment-neutron relative energy of 3~MeV.

\section{Models}
In this work, we consider predictions from three models. The first, the independent-particle shell model (IPSM), assumes that nuclear states are well described by one configuration, i.e., pure single-particle excitations.  While generally not a viable picture, the potential simplicity of configurations in $^{25,26}$F could allow it to be a reasonable first-order description of low-lying states.  Furthermore, all experimental considerations associated with assignments of $\ell$ values are done by comparison with this model. 

To gauge the validity of the IPSM picture in describing low-lying states in $^{25,26}$F, and to provide a more realistic account, we also compare to phenomenological shell-model calculations and ab initio valence-space IM-SRG.  For the former, we use the well-established USDA Hamiltonian \cite{Brown2006}, optimized to reproduce energy levels for all $sd$-shell nuclei. As in Ref. \cite{Lepailleur2015}, we choose USDA instead of USDB, since the latter is known to predict a too small excitation energy of the $J^{\pi} = 4^{+}$ state in $^{26}$F.
Valence-space IM-SRG has been shown to predict ground and excited states throughout the oxygen, fluorine, and neon isotopic chains \cite{Bogn14SM,Caceres2015,Stroberg2015,Stroberg2016}.  Beginning from nuclear forces derived from chiral effective field theory (EFT) \cite{Epel09RMP,Machleidt2011}, 3N forces between core and valence nucleons are typically captured by normal ordering with respect to the $^{16}$O reference. Without these initial 3N forces, the spectra of $^{25,26}$F are much too compressed, but with their addition, the IM-SRG spectra are in reasonable agreement with predictions from phenomenology \cite{Stroberg2015}. More generally, 3N forces are necessary to reproduce properties of exotic nuclei near oxygen and calcium \cite{Otsuka2010,Hebe15ARNPS,Hagen2012,Holt13Ox,Roth2012,Cipollone2013,Hergert2013PRL,Simonis2016,Holt2012,Hagen2012109,Gallant2012,Wienholtz2013,Soma2014,Hergert2014,Holt2014}. In this work we extend this approach to the ensemble normal-ordering procedure outlined in Ref.~\cite{Stroberg2016}, which accurately accounts for 3N forces between valence particles and reproduces results of large-space ab initio methods in all cases where those methods are reliable.  In all reported results, we use the same initial NN and 3N Hamiltonians as in Ref.~\cite{Stroberg2016}. To provide an uncertainty estimate from the many-body method, we perform calculations up to $e_{\mathrm{max}} \equiv 2n+\ell=14$ (with $n$ as the number of radial nodes, $\ell$ the orbital angular momentum, while $e_{\mathrm{max}} +1$ corresponds to the number of oscillator shells) and exponentially extrapolate to $e_{\mathrm{max}}=24$ for a range of harmonic-oscillator spacings $\hbar\omega\!=\!16-24$~MeV.  The resulting spread is indicated as a band in all IM-SRG results shown in Fig.~\ref{fig:LevelScheme25F} (right panel) and \ref{fig:LevelScheme26F} (right panel). In this approach, continuum effects are currently neglected.

Both USDA and IM-SRG calculations are performed within the standard $sd$ shell above an $^{16}$O core, and hence will only produce positive-parity states. In $^{25}$F, where we are also interested in exploring negative-parity states, we compare to predictions from the $s$-$p$-$sd$-$pf$ WBP interaction \cite{Warburton1990}, but with protons restricted to the $p$-$sd$ shells and neutrons restricted to $sd$ and at most two particles in each of the $f_{7/2}$, $p_{3/2}$, and $p_{1/2}$ orbits.  In all cases the shell-model diagonalization was carried out with the NushellX@MSU code \cite{Brown2014} to obtain the ground- and excited-state energies discussed below.

\section{Analysis and results} 
\subsection{Experimental results for $^{25}$F}
\label{par:25F}

\begin{figure}[htb]
\begin{center}
\includegraphics[width=8.7 cm]{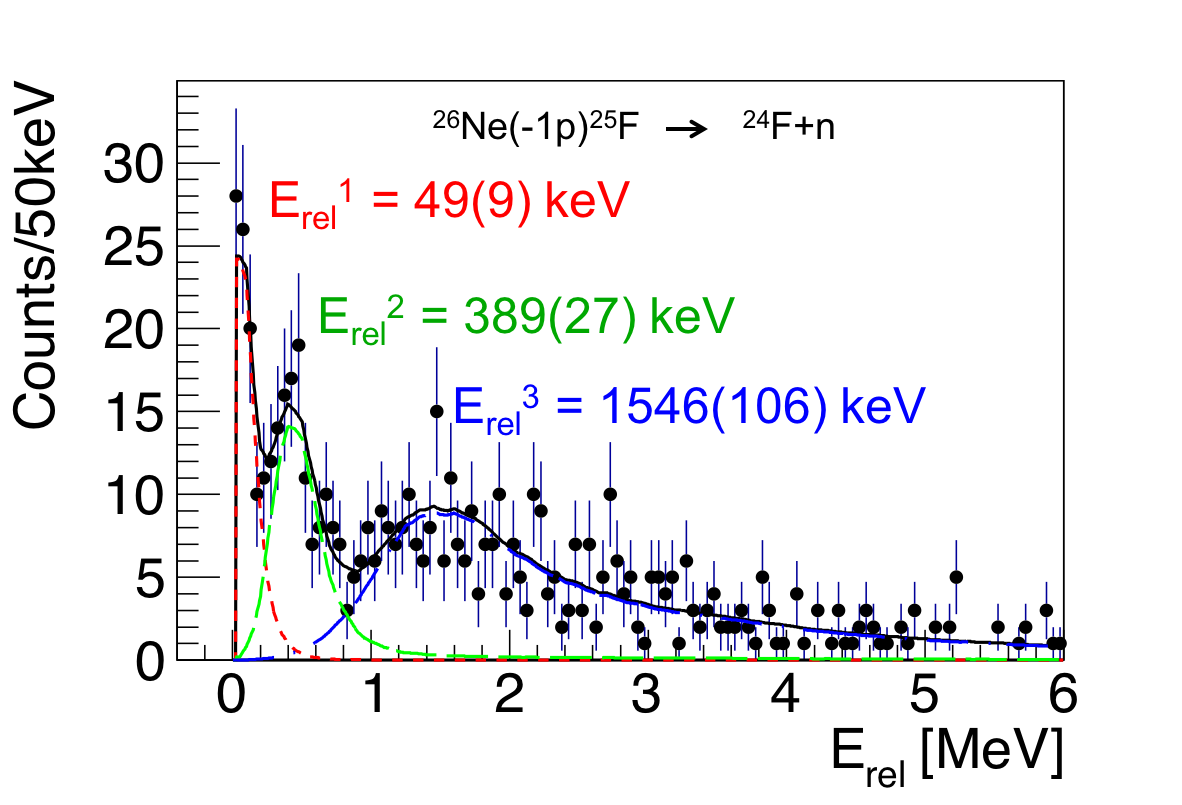}
\end{center}
\caption{(Color online) Relative energy spectrum of $^{25}$F. The solid black line shows the result of the fit composed of three resonances, marked in different colors, whose energies are written with uncertainties. Corresponding widths are given in Table~\ref{tab:25F}. In the fitting procedure, resonances were folded with the resolution of the LAND detector, enhancing their widths as compared to the intrinsic value.
\label{fig:estar25F}}
\end{figure}

The $^{25}$F nucleus was populated via the one-proton \mbox{($-$1p)} knockout reaction from a beam of $^{26}$Ne. To select this reaction channel, at least one proton must be detected in the Crystall Ball detector. This selection was made possible after the (neighboring-crystal) add-back treatment of the Crystal Ball data, in order to find the total deposited energy per particle, photon or proton, event by event. We note that two protons can be detected in the case of a (p,2p) reaction with the H nuclei of the target. The decay of unbound states, or resonances, of energies $E_\mathrm{r}^i$ in $^{25}$F will lead to the production of $^{24}$F, \textit{i)} either in its ground state, so the whole resonance energy $E_\mathrm{r}^i$ comes from the relative energy $E_\mathrm{rel}^i$ of the system ($^{24}$F+n), \textit{ii)} or in one of its excited states, that subsequently decays to the ground state by the emission of a $\gamma$ ray of energy $E_{\gamma}$. In the latter case, a coincidence between the neutron and the de-exciting $\gamma$ ray is observed. The excitation energies $E_\mathrm{exc}^i$ of the unbound states in $^{25}$F correspond to:
\begin{equation}
E_\mathrm{exc}^i = S_n + E_\mathrm{r}^i
          = S_n + E_\mathrm{rel}^i + (E_{\gamma})
\label{eq:Eexc}
\end{equation}
where $S_n$ is the neutron emission threshold. The relative energy was reconstructed on an event-by-event basis using the invariant mass equation with the momentum vectors of the fragment $^{24}$F and of the neutron:
\begin{equation}
\begin{split}
E_\mathrm{rel} = & \sqrt{m_{frag}^{2} + m_{n}^{2} + 2\left(\frac{E_{frag}E_{n}}{c^{4}} - \frac{p_{frag}p_{n}}{c^{2}}\cos \theta\right)}c^{2} \\
                 & - m_{frag}c^{2} - m_{n}c^{2} 
\end{split}
\end{equation}
In this equation,  $m_{frag}$ and $m_{n}$ are the rest masses of the fragment and the neutron, $E_{frag}$ and $E_{n}$ their total energies, $p_{frag}$ and $p_{n}$ their momenta, and $\theta$ their relative angle. As shown in Fig.~\ref{fig:estar25F}, the ($^{24}$F+n) relative energy spectrum displays three resonances. Each was described by a Breit-Wigner function whose width depends on its energy and on the orbital angular momentum $\ell$ of the emitted neutron \cite{Lane1958}:
\begin{equation}
\begin{split}
& f_{\ell}(E_\mathrm{rel}; E_\mathrm{rel}^i, \Gamma_\mathrm{r}) \\
&\propto \frac{\Gamma_{\ell}(E_\mathrm{rel})}{(E_\mathrm{rel}^i + \Delta_{\ell}(E_\mathrm{rel}) - E_\mathrm{rel})^{2} + \Gamma_{\ell}(E_\mathrm{rel})^2/4} 
\end{split}
\end{equation}
with the apparent width defined as:
\begin{equation}
\Gamma_{\ell}(E_\mathrm{rel}) = \Gamma_\mathrm{r} \times \frac{P_{\ell}(E_\mathrm{rel})}{P_{\ell}(E_\mathrm{rel}^i)} 
\end{equation}
and the energy shift given by:
\begin{equation}
\label{eq:shift}
\Delta_{\ell}(E_\mathrm{rel}) = \Gamma_\mathrm{r} \times \frac{S_{\ell}(E_\mathrm{rel}^i) - S_{\ell}(E_\mathrm{rel})}{2P_{\ell}(E_\mathrm{rel}^i)},
\end{equation}
where $P_{\ell}$ and $S_{\ell}$ are the penetrability and shift functions, respectively. This prescription, taken from Ref.~\cite{Hoffman2008}, ensures that the energy shift is eliminated at the resonance energy. The width, $\Gamma_\mathrm{r}$, extracted here in the one-proton knockout reaction is not corrected for the possible change of the overlap between the initial and final wave functions.

In order to extract the energy $E_\mathrm{r}^i$ and the intrinsic width $\Gamma_\mathrm{r}^i$ of the resonances, the spectrum of Fig.~\ref{fig:estar25F} is fitted, using the log likelihood method, with a linear combination of three $f_{\ell}$ functions which have been folded to include the resolution of the LAND detector, i.e. $\sigma \sim 260$~keV at $E_\mathrm{rel} \!= \!1$~MeV. In this fit, orbital angular momentum values $\ell$ between 0 and 2 were tested for each resonance, and the $\ell$ dependence is weak. A first resonance has been identified at $E_\mathrm{rel}^1 \!=\!  49(9)$~keV ($\Gamma_\mathrm{r}^1  \!=\! 51(49)$~keV), a second one at $E_\mathrm{rel}^2 \!=\! 389(27)$~keV ($\Gamma_\mathrm{r}^2  \!=\! 73(70)$~keV), and a third one at $E_\mathrm{rel}^3 \!=\! 1546(106)$~keV ($\Gamma_\mathrm{r}^3 \!=\! 2500(440)$~keV).
The extracted resonance centroids ($E_\mathrm{rel}^i$) and widths ($\Gamma_\mathrm{r}^i$) are based on the fit that uses ($f_{1}$, $f_{2}$, $f_{2}$) for the first, second and third resonances respectively, this combination will be justified in the section~\ref{SectionDiscussion25F}. The uncertainties correspond to one sigma and include only statistical error. 
For the $\ell=2$ resonances, the width of the resonances cannot be extracted easily due to the saturation of the Breit-Wigner line shape when the width $\Gamma_\mathrm{r}$ is increasing \cite{Hoffman2009}. To overcome this problem, another fit was performed in order to extract $\Gamma_\mathrm{r}$ for $\ell=2$ resonances. This fit uses simple Breit-Wigner functions without energy shift ($\Delta_{\ell}(E_\mathrm{rel}=0$)) and without angular dependence of the width ($\Gamma_{\ell}(E_\mathrm{rel})=\Gamma_\mathrm{r}$).
It should be noted that a non-resonant continuum has been estimated using the event-mixing procedure, which is based on the measured pairs of fragment + neutron \cite{Randisi2014}. This component was added as a free parameter in the fit. However, its contribution has been found to be negligible.

At this point, the relative energy spectrum and resonance energies of $^{25}$F ($E_\mathrm{rel}^i$) can be compared to those obtained previously at the National Superconducting Cyclotron Laboratory at Michigan State University \cite{Frank2011} using the same knockout reaction. While an almost continuous energy spectrum was obtained in \cite{Frank2011}, the better resolution achieved in the present work clearly allows us, despite the lower statistics, to distinguish at least three resonances. The energies of the three resonances proposed in Ref.~\cite{Frank2011}, ($E_\mathrm{rel}^1\!=\! 28(4)$~keV, $E_\mathrm{rel}^2 \simeq 350$~keV and $E_\mathrm{rel}^3 \simeq 1200$~keV), compare reasonably well with ours, considering the method-dependent determination of energy centroids in the case of broad resonances or for states lying very close to the neutron threshold. 
 
\begin{figure}[htb]
\begin{center}
\includegraphics[width=8.7 cm]{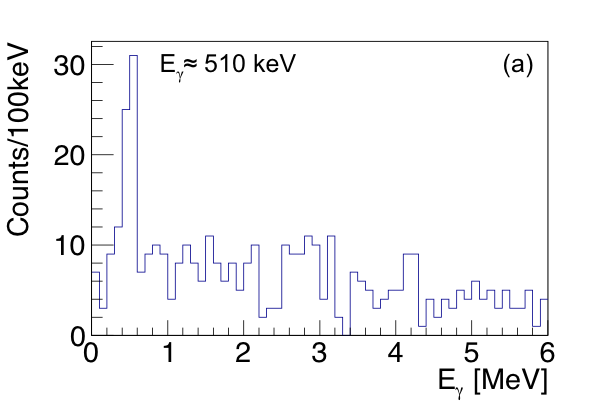}
\includegraphics[width=8.7 cm]{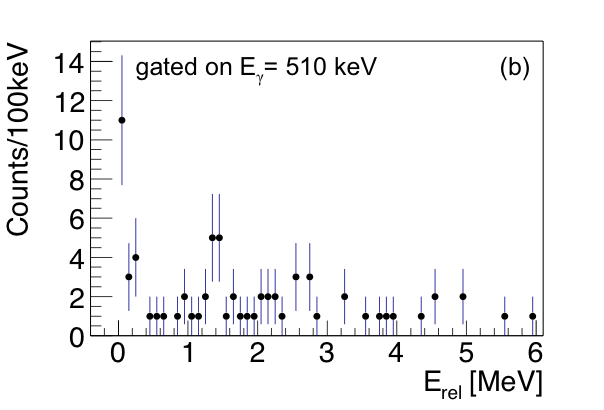}
\end{center} 
\caption{(Color online) (a) $\gamma$-ray spectrum obtained from the $^{26}$Ne(-1p) reaction, in coincidence with $^{24}$F and the detection of 
at least one neutron in LAND.  (b) Relative energy spectrum for $^{25}$F,  gated on the $\sim 510$~keV $\gamma$-ray transition in $^{24}$F.  
\label{fig:Gamma24F}}
\end{figure}

It is generally assumed, as in Ref.~\cite{Frank2011}, that unbound states decay with the largest available neutron energy. This means that the resonances would all decay to the ground state of $^{24}$F to maximize the $Q$-value of the neutrons. This assumption is often valid, except if the loss in $Q$-value when decaying to an excited state is compensated by a better matching between initial and final states. Neutron-$\gamma$ coincidences  are used to infer the energy of the resonances, $E_\mathrm{r}^i$, when the neutron decay proceeds to an excited state in $^{24}$F  followed by the emission of a $\gamma$ ray with energy $E_{\gamma}$ (Eq.~\ref{eq:Eexc}). Figure~\ref{fig:Gamma24F}(a) shows the presence of a peak near 510~keV in the neutron-gated $\gamma$-ray spectrum of $^{24}$F.  This peak likely corresponds to the decay of the $2^{+}_{1}$ to the $3^{+}$ ground state of $^{24}$F, observed at an energy of 521(1)~keV in the $\beta$ decay of $^{24}$O to $^{24}$F \cite{Caceres2015}.\footnote{We note that the 2\% energy difference observed between these two $\gamma$-rays is also observed in $^{26}$F: the decay of the $2^{+}_{1}$ to the $1^{+}$ ground state is observed, in our work, near 643~keV instead of 657(7)~keV in Ref. \cite{Stanoiu2012}.} Its presence in coincidence with neutron detection suggests that the decay of one or several resonances in $^{25}$F proceeds through this $2^{+}_{1}$ excited state, rather than directly to the ground state. 

The relative energy spectrum of Figure~\ref{fig:Gamma24F}(b), gated on the 510~keV $\gamma$-ray, displays a clear peak at the energy of the first neutron resonance, whose amplitude matches the one expected assuming a $\gamma$-ray efficiency of about 25\% and a 100\% branching to the $2^{+}_{1}$ excited state. It follows that the energy of the first resonance is $E_\mathrm{r}^1\!=\!49(9)+521(1)\!=\! 570(9)$~keV. No clear sign of $\gamma$ coincidence with the two other neutron resonances is observed. Indeed, from the amplitudes of the second and third resonances observed in the singles spectrum in Fig.~\ref{fig:estar25F},  approximately 40 and 100 neutrons should have been observed, respectively, in Fig.~\ref{fig:Gamma24F}(b) if these resonances  would have decayed 100\% to the $2^{+}_{1}$ excited state. The much lower number of observed neutrons, and the fact that no other $\gamma$-decay branches at higher energies than $521$~keV are observed in coincidence, indicates that the second resonance decays directly to the ground state of $^{24}$F. As a consequence, the second resonance is located at an excitation energy that is lower than the first one, in contradiction to the suggestion of Ref. \cite{Frank2011} where $\gamma$-coincidence information was not available. As for the third resonance, the few counts around 1400~keV in Fig.~\ref{fig:Gamma24F}(b) might be attributed to a coincidence with the $\gamma$ ray at 521(1)~keV. However, the marginal statistics, as well as the too narrow peak formed by these events, do not allow us to consider that the third resonance decays to the $2^{+}_{1}$ state in $^{24}$F. To summarize, excitation energies of $E_\mathrm{exc}^2=4659(104)$~keV, $E_\mathrm{exc}^1=4840(100)$~keV, and $E_\mathrm{exc}^3=5816(146)$~keV are deduced in $^{25}$F using Eq.~\ref{eq:Eexc}, with the relative energies $E_\mathrm{rel}^i$, the $\gamma$-coincidence information, and the neutron emission threshold of $S_n(^{25}\mathrm{F})\!=\!4270(100)$~keV \cite{Wang2012} (see Table~\ref{tab:25F}).

\subsection{Discussion on the results of $^{25}$F}
\label{SectionDiscussion25F}

\begin{table}[t]
\renewcommand{\arraystretch}{1.5}
 \begin{tabular}{  c c  c  c  c c  c }
    \hline \hline
    $i$ & $E_\mathrm{r}^i$ & $E_\mathrm{exc}^i$  & $\Gamma_\mathrm{r}^i$\ & $\Gamma_{sp}^{(\ell = 0)}$  & $\Gamma_{sp}^{(\ell = 1)}$ & $\Gamma_{sp}^{(\ell = 2)}$  \\
    \hline
    1 & $570(9)^{a}$ &  4840(100)	&  $51(49)$   &	1139	 	& 71 	 	& 0.5 \\
    2 &  $389(27)$    &  4659(104)      &  $73(70)$   & 3243		 & 1136	& 86\\
   3  & $1546(106)$ &  5816(146)	&  $2500(440)$& 6848	 & 4836	& 1799 \\
    \hline \hline
 \end{tabular}\\
 $^a$ The resonance located at the energy of $E_\mathrm{r}^1 \!= \!570(9)$~keV corresponds to the first peak in the relative energy spectrum of Fig.~\ref{fig:estar25F} at $E_\mathrm{rel}^1=49(9)$~keV, to which the coincident $\gamma$-ray energy of $E_{\gamma}=521(1)$~keV has been added (see text for details).
 \caption{\label{tab:25F} Characteristics of the $^{25}$F resonances   populated via one-proton-knockout $^{26}$Ne(-1p) under the assumption that $S_n\! = \!4270(100)$~keV \cite{Wang2012}. Resonance energies $E_\mathrm{r}^i$, excitation energies $E_\mathrm{exc}^i$, and widths $\Gamma_\mathrm{r}^i$ of the three resonances are given in keV with calculated single-particle widths $\Gamma_{sp}^{(\ell)}$, assuming various $\ell$ values of each resonance.}
\end{table}

\begin{figure*}
\begin{center}
\includegraphics[width=18 cm]{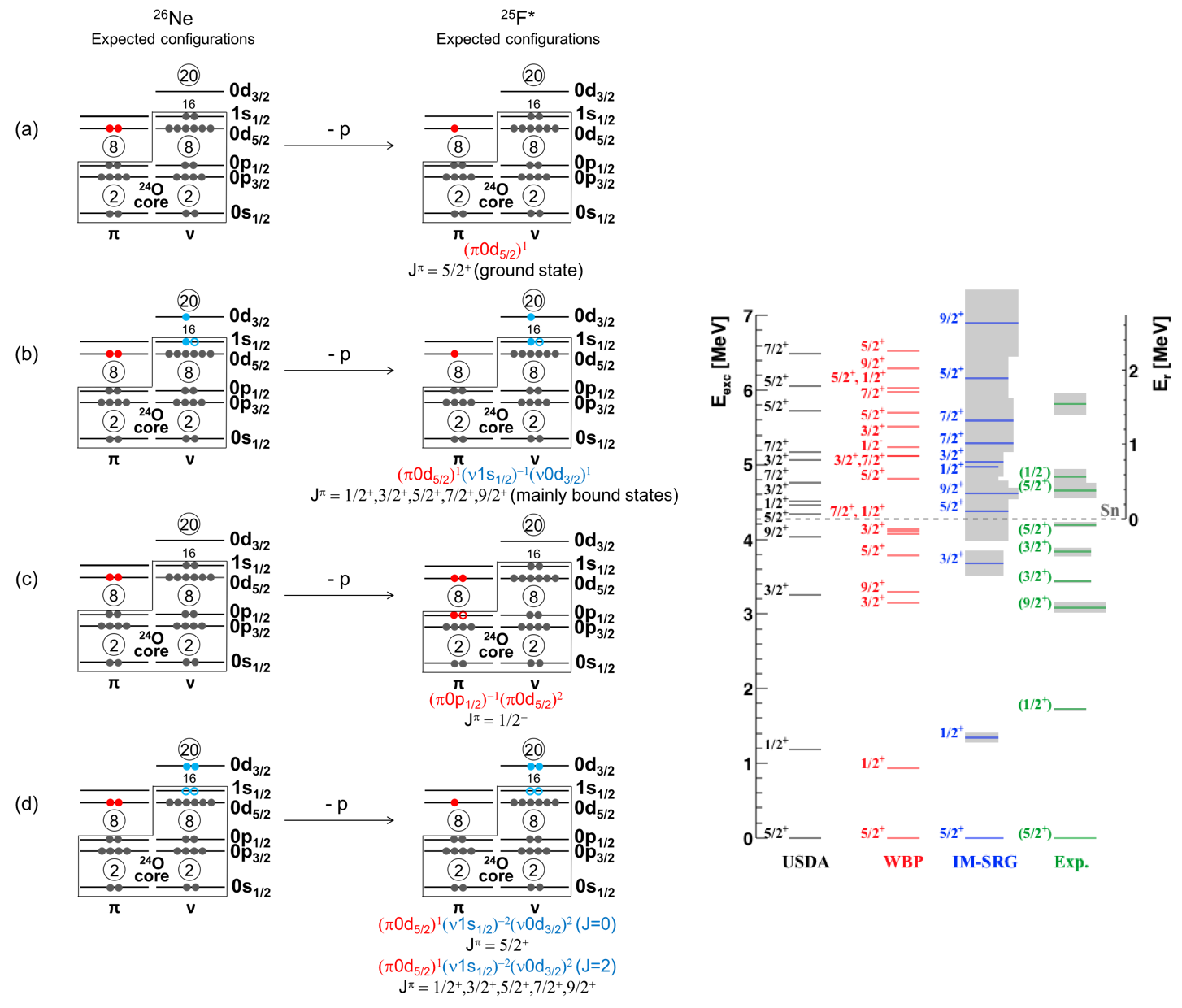}
\end{center}
\caption{(Color online) Left: Illustrative picture of the expected configurations populated in $^{25}$F from the $^{26}$Ne(-1p) reaction. Right: Experimental level scheme of $^{25}$F compared to shell-model calculations performed using the phenomenological USDA \cite{Brown2006} and WBP \cite{Warburton1990} interactions and with ab initio valence-space Hamiltonians derived from IM-SRG \cite{Stroberg2015,Stroberg2016}.  The unbound states above $S_n = 4270(100)$~keV were obtained in the present work, while the bound states were studied in Ref.~\cite{Vajta2014}. Grey rectangles shown in the experimental spectrum and in the IM-SRG predictions, with a $J^{\pi}$-dependent horizontal widths, correspond to uncertainties on the energy centroids of the states. These uncertainties on calculated energies overlap between 4 and 7~MeV, where many resonances are present. The bound states come mainly from $(\pi0d_{5/2})^1$ (case (a)) and $(\pi0d_{5/2})^{1}(\nu1s_{1/2})^{-1}(\nu0d_{3/2})^{1}$ (case (b)) configurations. We propose that unbound states come mainly from $(\pi0p_{1/2})^{-1}(\pi0d_{5/2})^{2}$ (case (c)) and $(\pi0d_{5/2})^{1}(\nu1s_{1/2})^{-2}(\nu0d_{3/2})^{2}$ (case (d)) configurations.
\label{fig:LevelScheme25F}}
\end{figure*}

In a simplified description of the $^{26}$Ne(-1p) reaction, protons are removed from the $\pi0 d_{5/2}$ orbit, leading primarily to the production of positive-parity (mostly bound) states in $^{25}$F (Fig.~\ref{fig:LevelScheme25F}(a,b)). If the $^{26}$Ne ground state contains some $2p2h$ neutron excitations $(\nu1s_{1/2})^{-2}(\nu0d_{3/2})^{2}$, positive parity states can also be  produced at higher excitation energy (likely above $S_n$) from a similar $\pi0d_{5/2}$ proton knockout (Fig.~\ref{fig:LevelScheme25F}(d)). Protons can be removed as well from the deeply bound  $\pi0p_{1/2}$ orbit, leading to negative parity states with mainly 1/2$^-$ spin-parity value (Fig.~\ref{fig:LevelScheme25F}(c)). We propose that the state at 4840~keV, which decays with a low energy of 49~keV to the excited state of $^{24}$F, rather than with a larger energy of 570~keV to the ground state, is a good candidate for a 1/2$^-$ state, since during the decay to the $2^{+}_1$ state in $^{24}$F, an $\ell\!=\!1$ neutron is emitted. A direct decay to the 3$^+$ ground state of $^{24}$F would imply that the neutron carried a larger angular momentum of $\ell\!=\!3$, which is strongly hindered.

It can be informative to compare experimental to calculated single-particle widths $\Gamma_{sp} (\ell)$ using various assumptions on $\ell$ values, ranging from 0 to 2. From this procedure, we would ideally  obtain further information on the nature and purity of each resonance. Using a Woods-Saxon potential whose depth is adjusted to reproduce the energy centroid of the resonance at 570(9)~keV, single particle widths of $\Gamma_{sp} ^{(\ell=0)} \!=\! 1139$~keV, $\Gamma_{sp}^{(\ell=1)} \!=\! 71$~keV and $\Gamma_{sp}^{(\ell=2)} \!=\! 0.5$~keV are calculated assuming pure configurations. The presently observed width of 51(49)~keV, is compatible with the $\ell\!=\!1$ assumption. This is in accordance with the earlier proposed $1/2^-$ spin-parity, derived from its observed decay to the 2$^{+}_1$ excited state of $^{24}$F.  As for the resonance at 389(27)~keV, among the calculated widths of $\Gamma_{sp}^{(\ell=0)}\!=\!3243$~keV, $\Gamma_{sp}^{(\ell=1)}\!=\!1136$~keV and $\Gamma_{sp}^{(\ell=2)}\!=\!86$~keV, a better agreement with the experimental width of $\Gamma_\mathrm{r}^2 = 73(70)$~keV is obtained with the $\ell\!=\!2$ configuration. This state would have a $5/2^+$ assignment if it corresponds to the configuration where a proton is knocked out from the $\pi0d_{5/2}$ orbit, with a neutron $(\nu1s_{1/2})^{-2}(\nu0d_{3/2})^{2}$ excitation (Fig.~\ref{fig:LevelScheme25F}(d)) in which neutrons are coupled to $J\!=\!0$. Other states (1/2$^+ - 9/2^+$) are considered when neutrons are coupled to $J\!=\!2$ (Fig.~\ref{fig:LevelScheme25F}(d)). 
With two neutrons in the $0d_{3/2}$ orbit, this tentative $5/2^+_3$ state likely decays through an $\ell\!=\!2$ neutron, leading to a final configuration $(\pi0d_{5/2})^{1}(\nu1s_{1/2})^{-2}(\nu0d_{3/2})^{1}$ in $^{24}$F. This coupling leads to  $J^{\pi} \!=\! 1^{+}_1 \!-\! 4^{+}_1$ states that were searched for by Caceres et al.~\cite{Caceres2015}. However, being at too high excitation energy, the $5/2^+_3$ state can only decay to the $J^{\pi}\!=\!3^{+}$ ground state  of $^{24}$F, that has a $(\pi0d_{5/2})^{1}(\nu1s_{1/2})^{1}$ configuration. It follows that the decay occurs through the low admixture of the $(\pi0d_{5/2})^{1}(\nu1s_{1/2})^{-2}(\nu0d_{3/2})^{1}$ component in the $J^{\pi}\!=\!3^{+}$ ground state. This feature implies an $\ell\!=\!2$ decay to the ground state with a low spectroscopic factor value.
For the resonance at 1546(106)~keV, the width $\Gamma_\mathrm{r}^3 \!=\! 2500(440)$~keV could correspond to several $\Gamma_{sp}^{(\ell=2)}$ states originating from the (1/2$^+ - 9/2^+$) multiplet (Fig.~\ref{fig:LevelScheme25F}(d)). It could alternatively correspond to another $\ell\!=\!1$ state. The characteristics of the identified resonances in $^{25}$F are summarized in Table \ref{tab:25F}. 

In Fig.~\ref{fig:LevelScheme25F} (right panel), we compare the measured experimental spectrum with the theoretical results. While both IM-SRG and USDA agree for a few excited states, the density of states given by USDA is higher than IM-SRG. The $5/2^+_2$ state predicted by both USDA and IM-SRG to be a $(\pi0d_{5/2})^1(\nu1s_{1/2})^{-1}(\nu0d_{3/2})^{1}$ configuration (Fig.~\ref{fig:LevelScheme25F}(b)) agrees well with the experimental state at 4.2~MeV.  The $5/2^+_3$ state from IM-SRG lies at  6.2~MeV but only contains a contribution from the neutron $2p2h$ configuration $(\nu1s_{1/2})^{-2}(\nu0d_{3/2})^{2}$ (Fig.~\ref{fig:LevelScheme25F}(d)) on the order of a few percent.  In USDA, the $5/2^+_3$ and $5/2^+_4$ states are close in energy at 5.7~MeV and 6.1~MeV, and both exhibit a very similar $2p2h$ character of approximately 30\%. We 
also note that the $J^{\pi}=1/2^{-}$ and $5/2^{+}_{3}$ states, calculated in Ref. 
\cite{Frank2011} at energies slightly above $S_n$ using the $psd$-shell 
WPM interaction are good candidates for the above resonances. By using  the WPB interaction mentioned in Section III, the newly proposed 1/2$^-$ resonance, that corresponds to a predominant proton cross-shell excitation, is calculated at 5.2~MeV, in reasonable agreement with experiment.

\subsection{Experimental results for $^{26}$F}
\begin{figure}[htb]
\begin{center}
\begin{minipage} {8 cm} 
\includegraphics[width=8.7 cm]{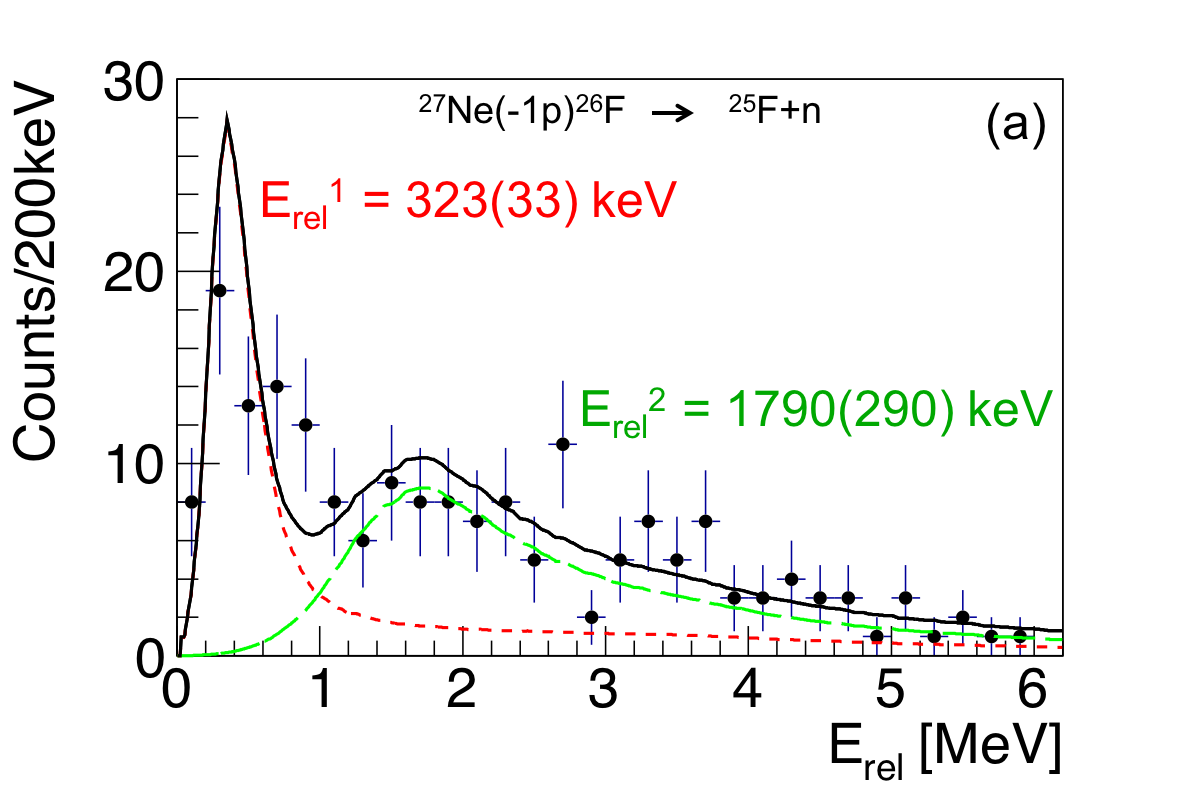}
\end{minipage}
\begin{minipage} {8 cm}
\includegraphics[width=8.7 cm]{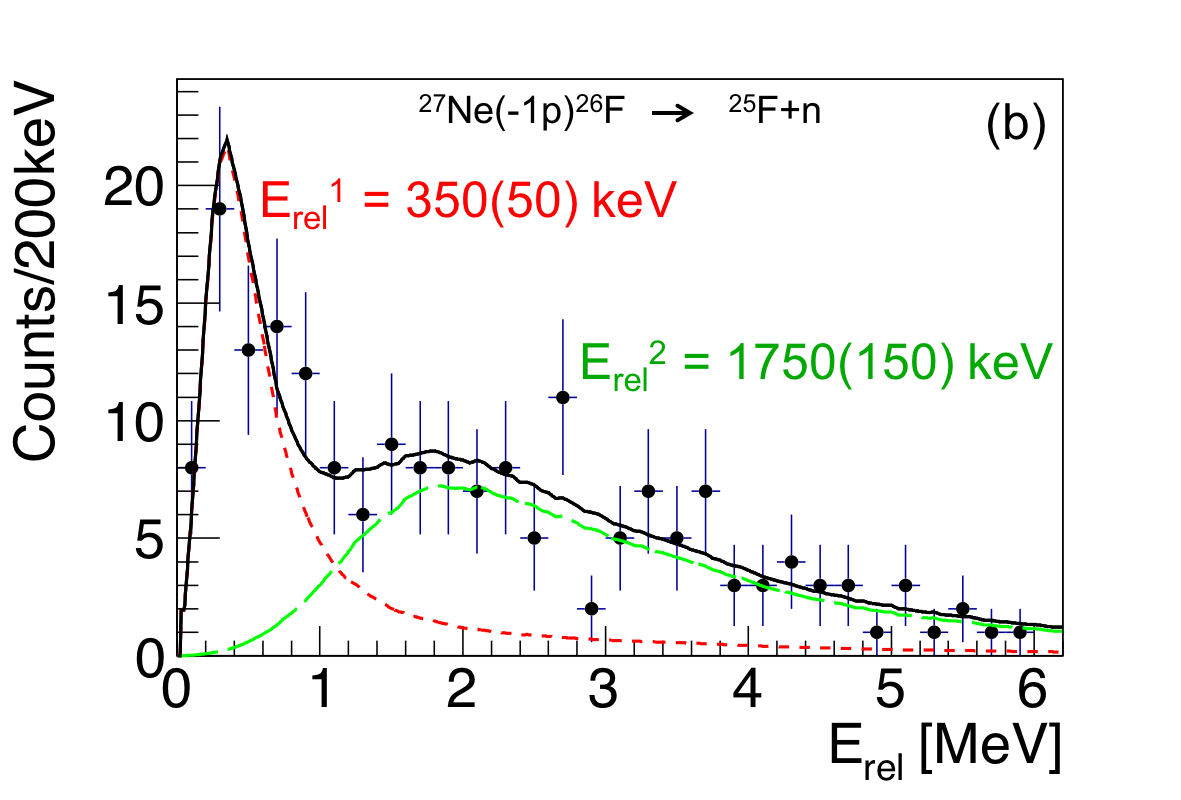}
\end{minipage}
\end{center}
\caption{(Color online) Relative energy spectrum for $^{26}$F. The solid black line shows the result of a Breit-Wigner fit using two $\ell \!=\! 2$ resonances whose centroid values are given in the figure. These resonances are folded with the resolution of the LAND detector. (a) With the energy shift $\Delta_{\ell}(E_\mathrm{rel})$ defined as in Eq.~\ref{eq:shift}. (b) Assuming $\Delta_{\ell}(E_\mathrm{rel}) = 0$.}
\label{fig:estar26F_delta}
\end{figure}

Unbound states of $^{26}$F were produced using the one-proton knockout reaction from $^{27}$Ne projectiles. The relative energy spectrum for $^{26}$F ($^{25}$F+n system), shown in Fig.~\ref{fig:estar26F_delta}(a), displays two resonances. No gamma is found in coincidence with them, implying that they decay directly to the ground state of $^{25}$F and that $E_\mathrm{rel}^i \!=\! E_\mathrm{r}^i$. Since the additional neutron in $^{26}$F likely occupies the $\nu0d_{3/2}$ orbital, an angular momentum $\ell\!=\!2$ has been used in the fit of each resonance, leading to $E_\mathrm{rel}^1 = 323(33)$~keV and $E_\mathrm{rel}^2 = 1790(290)$~keV. Pure $\ell\!=\!0$ resonances would be expected to be much broader, as shall be confirmed later in the discussion. The uncertainties correspond to one sigma confidence level. To check the sensitivity of the method, another fit was performed with a zero-energy shift in the $f_\ell$ function. In this way the extracted centroids of the resonances correspond to the maxima of the peaks \cite{Lecouey2002}. A  compatible result is found, with resonances at 350(50)~keV and 1750(150)~keV (Figure~\ref{fig:estar26F_delta}(b)). 

In order to determine the width $\Gamma_\mathrm{r}$ of these $\ell\!=\!2$ resonances, the same procedure as for $^{25}$F was applied (see paragraph~\ref{par:25F}).
$\Gamma_\mathrm{r}^1 \!= \!570(480)$~keV and $\Gamma_\mathrm{r}^2 \!= \!4200(2500)$~keV  were extracted for the first and second resonance widths. Owing to the fact that this fitting function is less adapted to the shape of the resonances, the errors bars on these energy centroids are larger, $E_\mathrm{rel}^1 \!= \!366(119)$~keV and $E_\mathrm{rel}^2 \!=\! 2430(650)$~keV, but the centroids themselves are fully compatible with those obtained previously and listed in Table~\ref{tab:26F}. 

Single-particle widths $\Gamma_{sp}^{(\ell =0)} \!=\! 3080$~keV, $\Gamma_{sp}^{(\ell =1)} \!=\! 1038$~keV, and $\Gamma_{sp}^{(\ell =2)} \!=\! 74$~keV are calculated for the first resonance at 323~keV (see Table~\ref{tab:26F}), which, within the large uncertainties, is compatible with an $\ell\!=\!2$ component. As for the second resonance at 1790~keV, $\Gamma_{sp} ^{(\ell =0)}\! =\! 7941$~keV, $\Gamma_{sp}^{(\ell =1)} \!=\! 6127$~keV and $\Gamma_{sp}^{(\ell =2)} \!=\! 2966$~keV are obtained. Considering the large uncertainty on the width of this resonance, it is difficult to conclude which $\ell$ assignment is preferred, and whether it corresponds to a single or to multiple overlapping resonances. The characteristics of the resonances identified in $^{26}$F, as well as the calculated single-particle widths, are summarized in Table~\ref{tab:26F}.

\begin{table}[t]
\renewcommand{\arraystretch}{1.5}
 \begin{tabular}{ c c  c  c  c  c  c }
    \hline \hline
    $i$  & $E_\mathrm{r}^i$ & $E_\mathrm{exc}^i$ & $\Gamma_\mathrm{r}^i$ & $\Gamma_{sp}^{(\ell = 0)}$ & $\Gamma_{sp}^{(\ell = 1)}$ & $\Gamma_{sp}^{(\ell = 2)}$  \\
    \hline
    1 & $323(33)$   &  1394(134)	 &   $570(480)$    & 	3080	 & 1038	 & 74   \\
    2  & $1790(290)$    & 2861(318) 	&  $4200(2500)$   & 	7941	&	6127	 & 2966\\
    \hline \hline
 \end{tabular}
 \caption{\label{tab:26F} Characteristics of the resonances measured in $^{26}$F populated via the one-proton knockout reaction $^{27}$Ne(-1p) under the assumption that $S_n = 1071(130)$~keV \cite{Jurado2007}. Resonance energies $E_\mathrm{r}^i$, excitation energies $E_\mathrm{exc}^i$, and widths $\Gamma_\mathrm{r}^i$ of the three resonances are given in keV with calculated single-particle widths $\Gamma_{sp}^{(\ell)}$, assuming various $\ell$ values of the resonance.}
\end{table}

\subsection{Discussion on the results of $^{26}$F}
\label{par:26FDiscussion}

\begin{figure*}
\begin{center}
\includegraphics[width=18 cm]{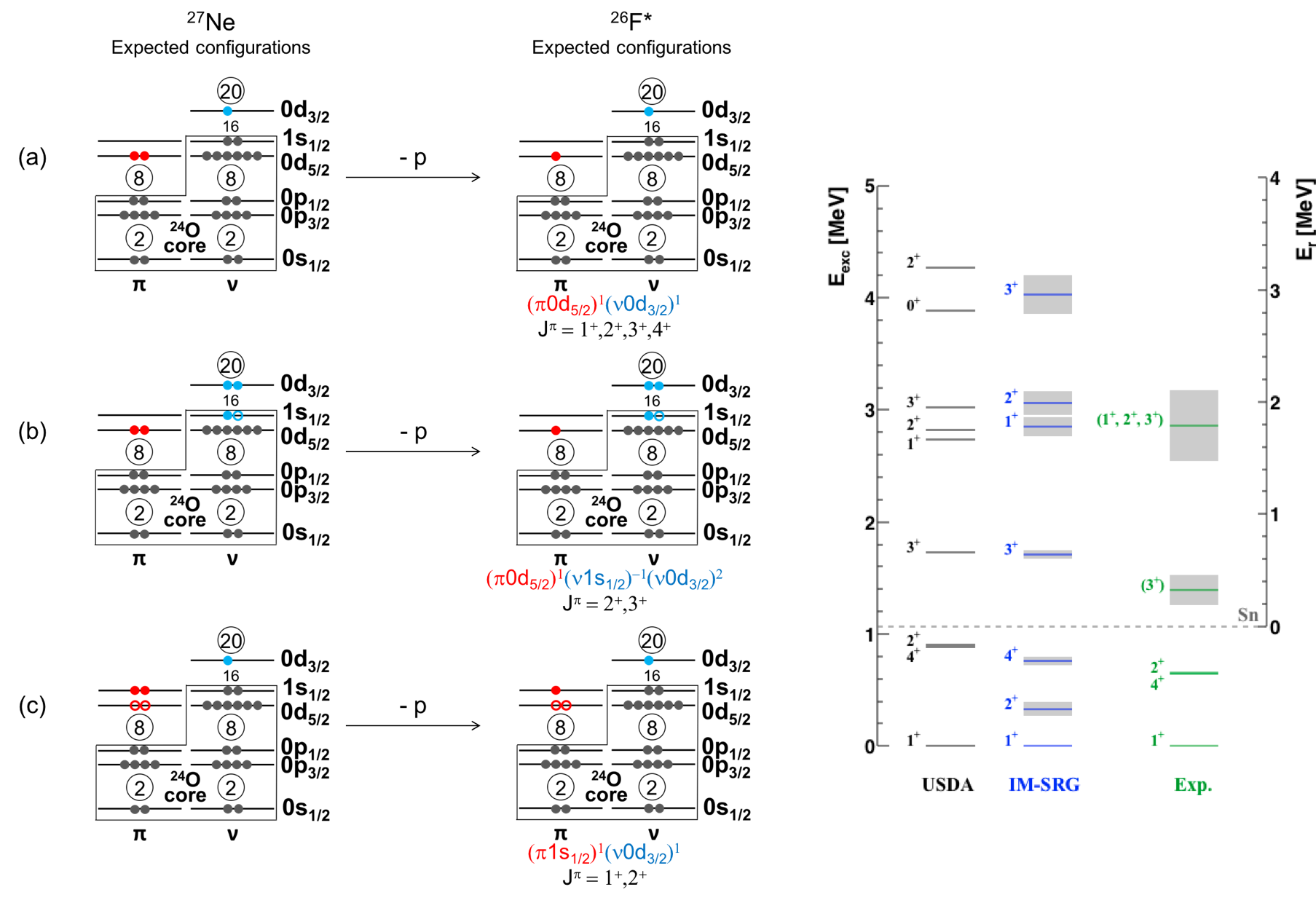}
\end{center}
\caption{(Color online) Left: Illustrative picture of the expected configurations populated in $^{26}$F from the $^{27}$Ne(-1p) reaction. Right: Experimental level scheme of $^{26}$F compared 
to shell-model calculations performed using the phenomenological USDA 
interaction \cite{Brown2006} and with ab initio valence-space Hamiltonians derived from IM-SRG \cite{Stroberg2015,Stroberg2016}. The energies of unbound states, above $S_n = 1071(130)$~keV, were newly 
measured in this work, while those of the bound states are taken from Refs.~\cite{Reed1999,Stanoiu2012,Lepailleur2013}. Grey rectangles shown in the experimental spectrum and in the IM-SRG predictions correspond to uncertainties on the energies centroids of the states and of the $S_n$. The bound states $J^{\pi} \!=\! 1^{+}_1,\! 2^{+}_1,\! 4^{+}_1$ as well as the unbound $J^{\pi}\!=\!3^{+}_1$ state are proposed to come from $(\pi0d_{5/2})^1(\nu0d_{3/2})^1$ (case (a)) configuration, while the second unbound state could come from $(\pi0d_{5/2})^{1}(\nu1s_{1/2})^{-1}(\nu0d_{3/2})^{2}$ (case (b)) and/or $(\pi1s_{1/2})^{1}(\nu0d_{3/2})^{1}$ (case (c)) configurations.
\label{fig:LevelScheme26F}}
\end{figure*}

It is reasonable to interpret low-lying states of $^{26}$F, which can be considered as one proton and one neutron outside a $^{24}$O core or one proton and three neutrons outside a $^{22}$O core, in terms of the IPSM. The simplest configuration would be $(\pi0d_{5/2})^{1}(\nu0d_{3/2})^1$ coupled above $^{24}$O, which generates the $J^{\pi} \!=\! 1^{+}_1 \!-\! 4^{+}_1$ multiplet (Fig.~\ref{fig:LevelScheme26F}(a)). Among these states, only the $J\!=\!3^{+}_{1}$ has not been experimentally observed. The next most likely multiplets in the IPSM arise from the $(\pi0d_{5/2})^1(\nu1s_{1/2})^{-1}(\nu0d_{3/2})^2$ configuration above $^{24}$O, leading to a $J^{\pi} \!=\! 2^{+},  3^{+}$ doublet (Fig.~\ref{fig:LevelScheme26F}(b)), and the $(\pi1s_{1/2})^1(\nu0d_{3/2})^1$ configuration above $^{24}$O, leading to a  $J^{\pi}\!=\! 1^{+},  2^{+}$ doublet (Fig.~\ref{fig:LevelScheme26F}(c)).  

A resonance $271(37)$~keV above the neutron threshold  
was previously observed in $^{26}$F using the nucleon-exchange reaction 
$^{26}$Ne $\rightarrow$ $^{26}$F \cite{Frank2011}, in which one  
$\pi0d_{5/2}$ proton in $^{26}$Ne is converted into a neutron in the 
$\nu0d_{3/2}$ orbital, and should produce all states of the 
$J^{\pi} \!=\! 1^{+}_1 \!-\! 4^{+}_1$ multiplet. As favored by this reaction mechanism, the $271(37)$~keV 
resonance could correspond to the missing $J^{\pi}\!= 3^{+}_1$ state of 
the multiplet, but no spin assignment was proposed in 
Ref.~\cite{Frank2011}. The knockout of a $\pi0d_{5/2}$ proton 
from $^{27}$Ne will also leave the $^{26}$F nucleus in a similar 
$(\pi0d_{5/2})^1(\nu0d_{3/2})^1$ configuration and produce the same 
multiplet of states (Fig.~\ref{fig:LevelScheme26F}(a)). Therefore, the fact that the {\it same} resonance is 
observed in the two experiments, at 271(37) keV in Ref.~\cite{Frank2011} and at 
323(33)~keV in the present work, gives further confidence in the 
assignment of this resonance as a $J^{\pi} \!=\! 3^{+}_1$ state. The width of 
the resonance, in accordance with an $\ell\!=\!2$ emission, also supports 
this assignment.

No other resonance was observed in Ref.~\cite{Frank2011}. In the knockout reaction, 
higher energy resonances would be produced only when some neutron or proton admixture is present in the $^{27}$Ne ground state. 
Results of the $^{26}$Ne$(d,p)^{27}$Ne \cite{SMBrown2012} 
transfer reaction have revealed that some neutron excitations across 
$N\!=\!16$ occur, i.e., $(\nu1s_{1/2})^{-1}(\nu0d_{3/2})^2$, as indicated by 
the partial vacancy of the $\nu1s_{1/2}$ orbit and the increased occupancy of 
the $\nu0d_{3/2}$ orbit. This offers the possibility to produce the 
$J^{\pi}\!=\! 2^{+}, \, 3^{+}$ states in $^{26}$F from the knockout of 
$^{27}$Ne (Fig.~\ref{fig:LevelScheme26F}(b)), making the second (broad) resonance a good candidate for one or two of these states.

Proton $(\pi0d_{5/2}$)$^{-2}(\pi1s_{1/2})^{2}$ admixtures in the ground 
state configuration of $^{27}$Ne are also possible (Fig.~\ref{fig:LevelScheme26F}(c)), as the two proton orbits are relatively close in energy (a 1/2$^+$ state, originating from the $(\pi0d_{5/2}$)$^{-2}(\pi1s_{1/2})^{1}$ configuration, has been proposed 
at 1720(15)~keV in $^{25}$F \cite{Vajta2014}). This would produce $J^{\pi}\!=\! 1^{+}, \, 2^{+}$ resonances in $^{26}$F, populated in the 
knockout reaction from $^{27}$Ne. While not excluded, such $2p2h$ proton excitations in $^{27}$Ne are unlikely for two reasons. First, the pairing energy, which scales with ($2j+1$), in principle favors keeping protons in the $\pi0d_{5/2}$ orbit rather than promoting them to the upper $\pi1s_{1/2}$ orbit. Second, from the analysis of the one-proton knockout reaction in $^{26}$Ne, we find an upper value of 8\% for the direct feeding of the 1/2$^+$ state at 1720(15)~keV in $^{25}$F and therefore for the $2p2h$ content of the $^{26}$Ne ground state.  If the two $2^{+}$ states (of the $J^{\pi}\!=\! 2^{+}, \, 3^{+}$ and the $J^{\pi}\!=\! 1^{+}, \, 2^{+}$ multiplets) were produced from these neutron or proton excitations, their configuration would likely be mixed, especially if the resonances lie close in energy. Comparison with shell-model calculations will now help complete this qualitative discussion.

In Fig.~\ref{fig:LevelScheme26F} (right panel), the proposed experimental level scheme for $^{26}$F is compared to results of phenomenological shell-model calculations from the USDA Hamiltonian and ab initio valence-space IM-SRG. Both calculations reproduce the energies of the 
two bound excited states in $^{26}$F,  $J^{\pi} = 2^+_1$ and 
$4^+_1$. In addition the one neutron separation energy predicted by IM-SRG of $S_n\!=\!1020(100)$~keV agrees well with experiment. The first excited state above the neutron threshold 
likely corresponds to the $J^{\pi} = 3^{+}_1$ state belonging to the 
$J^{\pi} = 1^+_1 - 4^+_1$ multiplet, lying within several hundred keV of both USDA and IM-SRG predictions.  Its calculated neutron 
occupancies, which are approximately 1.9 $\nu1s_{1/2}$ and 1.3
$\nu0d_{3/2}$, for both IM-SRG and USDA, correspond to a predominant 
($\nu1s_{1/2})^2(\nu0d_{3/2})^1$ single-particle configuration (Fig.~\ref{fig:LevelScheme26F}(a)).  Moreover, these 
occupancies are nearly identical to those of all other members of the 
$J^{\pi} = 1^+_1 - 4^+_1$ multiplet for both interactions.
The calculations also predict unbound $J^{\pi} \! = \! 2^+_{2,3}$, $1^+_2$, and $3^+_2$ states at higher excitation energies, with occupancies corresponding to the IPSM configurations estimated earlier. From IM-SRG and USDA, the $J^{\pi} \!=\! 2^+_3, 3^{+}_2$ states have  approximately a 1.9 and 2.0 $\nu0d_{3/2}$ occupancy, respectively, compatible with a $2p1h$ excitation.  
With an occupancy of 0.8 in the $\pi1s_{1/2}$  orbital, both calculations predict the $J^{\pi} = 1^{+}_2$ state to correspond to the proton excitation configuration of Fig.~\ref{fig:LevelScheme26F}(c). The $J^{\pi} = 2^{+}_2$, however, has a more mixed configuration 
between a proton excitation to the $\pi1s_{1/2}$ orbital and a neutron promoted to the 
$\nu0d_{3/2}$ orbital. Experimentally, only a broad resonance, centered at 
about 1790~keV, is observed. This broad component can encompass the 
three lowest calculated resonances  $2^{+}_2, 1^{+}_2$ and $3^{+}_2$ 
that lie within 1~MeV of excitation energy. Despite this general agreement, we note a systematic shift of several hundred keV between IM-SRG predictions and the experimental resonances that probably arise from the fact that the IM-SRG calculations use harmonic oscillator basis and treat unbound states as if they were bound.

One important word of caution must be added concerning the $S_n$ 
value of $^{26}$F and its consequence on a possible shift in excitation 
energy of the resonances. The tabulated $S_n$ value of $0.80(12)$~MeV was derived from a time-of-flight measurement of  
$^{26}$F nuclei produced in a fragmentation reaction \cite{Jurado2007} in 
which the existence of the 4$^{+}_1$ isomer at 643~keV 
\cite{Lepailleur2013} was not known. Therefore, this value possibly contains some mixture of the ground state and isomeric states, and should be considered as a lower value of $S_n$. By assuming an isomeric ratio of 
42(8)\%, derived from the production of $^{26}$F in 
the same fragmentation reaction \cite{Lepailleur2013}, the $S_n$ value is increased by 
270(50)~keV, yielding $S_n= 1.07(13)$~MeV. This corresponds to the  
$S_n$ value adopted in Fig.~\ref{fig:LevelScheme26F} (right panel). If the isomeric 
ratio were 100\%, the $S_n$ value  would reach $1.44(12)$~MeV, and the 
excitation energy of all resonances would increase by 373~keV,  bringing 
the 3$^{+}_1$ closer to the USDA and IM-SRG theoretical predictions.

We now turn to experimental interaction energies, Int$(J)^{\mathrm{exp}}$, which in the IPSM limit would correspond to the interaction between a $0d_{5/2}$ proton and a $0d_{3/2}$ neutron above a $^{24}$O core coupled to different spin orientations $J$.  We define this quantity in terms of the experimental energies in $^{25,26}$F, $^{24}$O and $^{25}$O following the formalism  of Ref.~\cite{Lepailleur2013}:
\begin{equation} \label{intexp}
\mathrm{Int}(J)=\mathrm{BE}(^{26}\mathrm{F})_J-\mathrm{BE}(^{26}\mathrm{F})_{\mathrm{free}}, 
\end{equation}
where 
\begin{equation} \label{BEfree}
\mathrm{BE}(^{26}\mathrm{F})_{\mathrm{free}}=\mathrm{BE}(^{25}\mathrm{F})+\mathrm{BE}(^{25}\mathrm{O})-\mathrm{BE}(^{24}\mathrm{O}),
\end{equation}
and BE($^{26}$F)$_J$ is the energy of a given $J^{\pi}$ state in $^{26}$F.
Values of Int(1,2,4)$^{\mathrm{exp}}$, obtained in Ref.~\cite{Lepailleur2013}, and \mbox{$\mathrm{Int}(3)^{\mathrm{exp}}= -0.45(19)$~MeV}, derived from the $3^+_1$ energy measured in the present work, are listed in Table 
\ref{int26F}. The corresponding \emph{effective experimental} monopole interaction (i.e., 
the $J$-averaged interaction energy) amounts to 
\mbox{$V_{\mathrm{pn}}^{\mathrm{exp}} \simeq -1$ MeV}.

\begin{table}[t]
\renewcommand{\arraystretch}{1.4}
 \begin{tabular}{ c | c  c  c  c  c |  c }
 \hline \hline

     \multirow{2}*{$J$} & 	 		\multicolumn{5}{c|}{Int($J$) [MeV]} 		&  \multirow{2}*{$R(J) $[\%]} \\
        & exp & $\delta$ & $\delta + corr $ & IM-SRG & USDA & \\
    \hline
    1 & -1.85(13)$^a$ & -1.85$^b$	&	-1.85$^b$	&-2.24(07) 	& -2.47 	& 100$^b$    \\
    2  & -1.19(14)    	& -0.90	   	& -0.82		& -1.86(05) 	& -1.51	&  91 \\
    3  & -0.45(19)    	& -0.37 	   	& -0.28		& -0.53(04) 	& -0.69	&  74 \\
    4  & -1.21(13)    	& -1.32 	  	& -1.21		& -1.56(04) 	& -1.54	&  91  \\
    \hline
    $V_{\mathrm{pn}}$ & -1.06(8) & -1.02 & -0.94 	& -1.41(02) 	& -1.40	&  \\
    \hline \hline

 \end{tabular}\\
$^a$ Obtained when using $S_n = 1.07(13)$~MeV\cite{Jurado2007,Lepailleur2013}.\\
 $^b$ Normalized to Int(1)$^{\mathrm{exp}}$ .\\
 \caption{\label{int26F} Experimental and  calculated interaction energies, Int$(J)$ in MeV, between a $0d_{5/2}$ proton and a $0d_{3/2}$ neutron in $^{26}$F.  Calculated results are obtained from USDA and IM-SRG shell model calculations and a schematic $\delta$ interaction. $R(J)$ denotes a correction applied to Int$(J)^{\delta}$ to deduce the interaction energy Int$(J)^{\delta+corr}$ (see text for details).}
\end{table}

\begin{figure*}
\begin{center}
\includegraphics[width=18 cm]{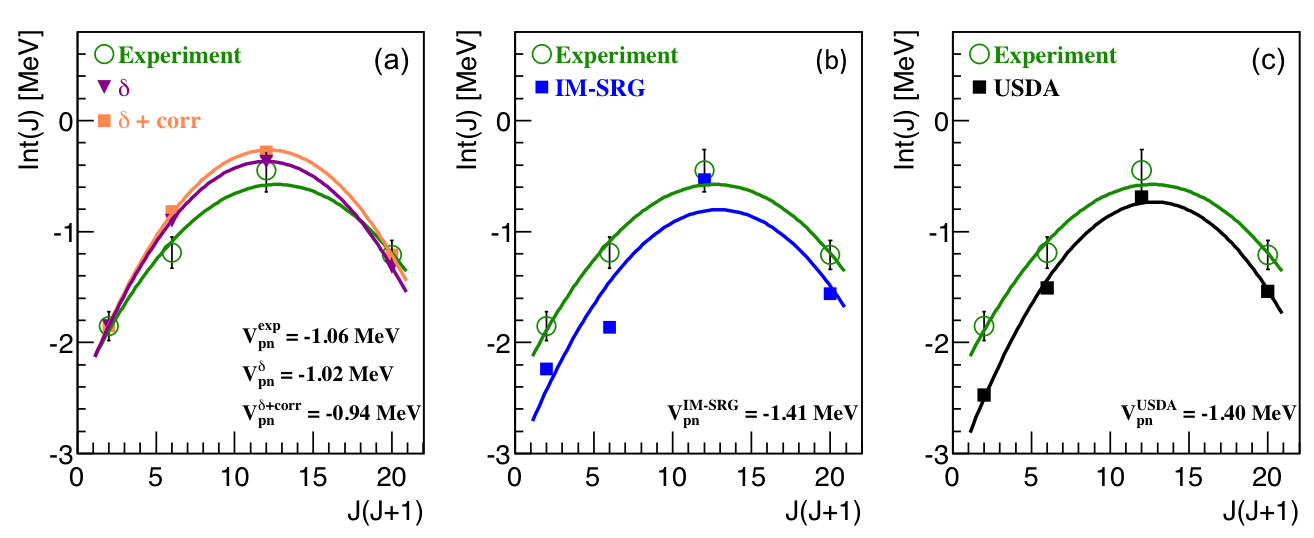}
\end{center}
\caption{Experimental interaction energies corresponding to the $\pi 0d_{5/2} \times \nu 0d_{3/2}$ coupling in $^{26}$F, Int$(J)^{\mathrm{exp}}$ (green circles), are plotted as a function of $J(J+1)$ and compared to calculations using (a) a delta interaction without (\color{violet}$\blacktriangledown$\color{black}) or with (\color{orange}$\blacksquare$\color{black}) $J$-dependent radial corrections (see text for details), (b) the IM-SRG procedure, and (c) the USDA interaction. Fitted parabolas are drawn to guide the eye. Extracted experimental and calculated monopole values $V_{pn}$ are given in each panel. All values of Int$(J)$ and $V_{pn}$ are given in Table \ref{int26F}.}
\label{fig:int26F}
\end{figure*}

For comparison, we first consider Int$(J)^\delta$, calculated in a simple picture of a 
proton-neutron system interacting via a zero-range 
$\delta$-interaction, decomposed into radial 
$F_R(n_p,\ell_p,n_n,\ell_n)$ and angular $A(j_p,j_n,J)$ parts \cite{Heyde}: 
\begin{equation}
\mathrm{Int}^\delta(j_p,j_n,J)= F_R(n_p,\ell_p,n_n,\ell_n) A(j_p,j_n,J),
\end{equation}
where the radial overlap between the proton and neutron wave functions 
is:
\begin{equation}\label{radial}
F_R(n_p,\ell_p,n_n,\ell_n)= \frac{V_0}{4\pi} \int_0^\infty \frac{1}{r^2}[R_{n_p,\ell_p}(r)R_{n_n,\ell_n}(r)]^2dr
\end{equation}

We account for the unknown strength of the nuclear interaction $V_0$, by normalizing Int$^\delta(J)$ to experimental data, i.e.\ in the present case to Int$(1)^{\mathrm{exp}}$. The angular part, $A(j_p,j_n,J)$, lifts the degeneracy between the different $J$ states of the multiplet and is independant of the choice of the nuclear interaction. In Fig.~\ref{fig:int26F}, the values of Int$(J)^{\mathrm{exp}}$ display an upward parabola as a function of $J(J+1)$. As expected, Int$(1)^{\mathrm{exp}}$ and Int$(4)^{\mathrm{exp}}$, that correspond to the coupling of a proton in $0d_{5/2}$ and a neutron in $0d_{3/2}$ in coplanar orbits, have the strongest intensities.

Contrary to well bound systems, the radial overlap between the proton and the neutron becomes poorer from one $J$ state to another as the neutron becomes less and less bound. This introduces an implicit $J$-dependence of the radial part $F_R(n_p,\ell_p,n_n,\ell_n)$ that we shall characterize by a reduction factor $R(J)$. To determine $R(J)$, the proton-neutron radial overlap was calculated using experimental neutron binding energies in the $^{26}$F system. The corresponding wave functions were obtained by solving the Schr\"{o}dinger equation in a Woods-Saxon potential, with a depth adjusted to reproduce the observed neutron or proton binding energies for the states of the $^{26}$F multiplet. Compared to the $1^+$ state, smaller radial overlaps are found for other $J$ states, which we characterize with the reduction factors, $R(J)$, shown in Table \ref{int26F}. Being the least bound, the $J=3$ state experiences the largest correction factor $R(J)$ of 74\%. Applying this $J$-dependent correction $R(J)$ on the radial wave function, that leads to Int$(J)^{\delta+corr}$. Despite the largest reduction factor $R(J)$ for the $J=3$ state, the Int$(3)^{\delta}$ value of $-0.37$~MeV is only slightly modified by about 100~keV (Int$(3)^{\delta+corr} = -0.28$~MeV) owing to its weak intensity. As shown in Table \ref{int26F}, both calculated interaction energies, Int$(J)^{\delta}$ and Int$(J)^{\delta+corr}$, compare reasonably well with experimental values, Int$(J)^{\mathrm{exp}}$.  This shows that a fairly good description of the amplitude of the multiplet is obtained with this schematic model, with a modest shift of the unbound $J=3$ state as compared to if it was treated as a bound state.

We add for comparison in Table \ref{int26F} and Fig.~\ref{fig:int26F}  interaction energies obtained from the USDA and IM-SRG calculations using equations \ref{intexp} and \ref{BEfree}. This way, experimental and theoretical Int$(J)$ are directly comparable, since they include correlations on equal footing. For USDA and IM-SRG the monopole interaction $V_{pn}$ amounts to about $-1.4$~MeV. This is larger than the experimental value of $-1.06$~MeV, pointing to a smaller monopole interaction as compared to calculations. As seen in Table \ref{int26F} and in Fig.~\ref{fig:int26F}, the amplitude of the multiplet parabola of USDA is also larger than experiment, while the energy of the $J=3$ state is in good agreement. 
This suggests that the residual energy, that lifts the degeneracy between the J-components of the multiplet, is smaller than calculated. Both effects of smaller monopole and residual interactions, as compared to calculations, could be interpreted (with the word of caution concerning the binding energy of the $^{26}$F ground state mentioned before) as an effect of the proximity of the continuum on the effective proton-neutron interaction. 
We note that the IM-SRG values, \emph{not} normalized to any experimental data, reproduce the Int$(J)^{\mathrm{exp}}$ values, though with some overbinding. This is likely due to the starting SRG-evolved NN+3N Hamiltonians, which are known to gradually overbind with increasing nucleon number past $^{16}$O \cite{Stroberg2015,Stroberg2016}. 

\section{Conclusion}

Unbound states in $^{25,26}$F have been studied using the one-proton 
knockout reaction from $^{26,27}$Ne projectiles. Resonances at 
49(9)~keV, 389(27)~keV, and 1546(106)~keV were measured in $^{25}$F. 
Being in coincidence with the 521~keV $\gamma$ transition, the energy of the  $2^+_1\rightarrow$ g.s. transition in $^{24}$F, the energy of the 
first resonance must be shifted upward compared to the value derived in Ref.~\cite{Frank2011}, where $\gamma$-ray detection was not available. This state at $E_\mathrm{exc}^1 = 4840(100)$~keV is a good candidate for a proton $\pi0p_{1/2}$ hole $(1/2^-$ state) configuration, as discussed in comparison to shell-model calculations using the WBP interaction. 

Unbound states in $^{26}$F have been studied using the same procedure. Two resonances have been observed at 323(33)~keV and 1790(290)~keV. The first resonance  has been identified as a convincing candidate for the $3^+_1$ state of the $J^{\pi} = 1^{+}_1 - 4^{+}_1$ multiplet, based on its observation in the two selective reactions of charge exchange from $^{26}$Ne and of knockout from $^{27}$Ne, as well as its relatively narrow width pointing to $\ell = 2$ neutron configuration. The second broad resonance, not observed in previous studies, might reflect several states that could not be distinguished, corresponding to neutron $(2p1h)$ or proton $(1p)$ components. 

These  $J^{\pi} = 1^{+}_1 - 4^{+}_1$ states, arising from the ($\pi0d_{5/2})^1(\nu0d_{3/2})^1$ coupling, are particularly adapted to probe the evolution of Int$(J)$ close to the neutron drip line. A resulting effective interaction \mbox{$V_{\mathrm{pn}}^{\mathrm{exp}} \simeq -1$ MeV} has been found for this proton in the $0d_{5/2}$ orbital and this neutron in the $0d_{3/2}$ orbital. Energies of these $J^{\pi} = 1^{+}_1 - 4^{+}_1$ states have been compared with phenomenological shell-model calculations using the USDA interaction and ab initio valence-space IM-SRG calculations. In the two cases, an overall good agreement between predicted and measured energies is found for the bound states. However, higher-lying states are found to be too high in energy, highlighting the need to include coupling to continuum in the models for broad resonances. It is deduced here that, as compared to models that use an harmonic oscillator basis to determine the wave functions of the nucleons independently of their binding energy, (i) the overall effective interaction is weakened by about 30-40\% and (ii) the amplitude of the multiplet of $J^{\pi} = 1^{+}_1 - 4^{+}_1$ states is more compressed, though correlations (overlap between the $0d_{5/2}$ proton and the $0d_{3/2}$ neutron wave functions) are still strong enough to lift the degeneracy between these $J$ states. 

To summarize, as shown in this paper and in references~\cite{Stanoiu2012,Lepailleur2013,Lepailleur2015,Frank2011,Stefan2014}, $^{26}$F, which is close to the doubly magic $^{24}$O nucleus, is particularly adapted to study the effects of the coupling to continuum through the changes in binding energy and the width of its unbound states. These studies provide stringent constraints for future theoretical development including the treatment of the continuum and aiming at a better description of shell evolution at the drip lines.
In the future, the increased granularity of the neutron detectors, as well as a longer time-of-flight basis, will lead to a better energy resolution. This will allow to disentangle overlapping resonances, herewith providing access to their width and to their coupling to bound or unbound states. We finally note that a large part of the conclusions drawn here rely on the $S_n$ value of $^{26}$F that is subject to uncertainties because its atomic  mass was measured with an unknown fraction of the $J=4^+$ isomer at 643~keV. We therefore strongly encourage to confirm the $S_n$ value of $^{26}$F to put the comparison between experiment and theory on a more reliable basis.

\begin{acknowledgments}
P. Van Isacker and M. Ploszajczak are greatly acknowledged for fruitful discussions and suggestions on how to improve the manuscript.
TRIUMF receives funding via a contribution through the National Research 
Council of Canada. This work was supported in part by NSERC, the 
NUCLEI SciDAC Collaboration under the U. S. Department of Energy 
Grants No. DE-SC0008533 and DE-SC0008511, the National Science 
Foundation under Grants No.PHY-1404159, the European Research 
Council Grant No.307986 STRONGINT, the Deutsche Forschungsgesellschaft under Grant SFB 1245, and the BMBF under Contracts 
No. 05P15RDFN1 and 05P15WOFNA. This work has also been supported by the Spanish MINECO via project FA2013-41267-P, FPA2015-64969-P and by the Portuguese FCT, Project PTDC/FIS/103902/2008. Computations were performed with an 
allocation of computing resources at the J\"ulich Supercomputing Center, 
Ohio Supercomputer Center (OSC), and the Michigan State University 
High Performance Computing Center (HPCC)/Institute for Cyber-Enabled 
Research (iCER). C. A. Bertulani acknowledges support from U. S. DOE Grant DE-FG02-08ER41533 and the U. S. NSF Grant No.1415656. M. Petri acknowledges support from the Helmholtz International Center for FAIR
within the framework of the LOEWE program launched by the State of Hesse.
\end{acknowledgments}

\bibliographystyle{myapsrev4-1}
\bibliography{paper}

\end{document}